\documentclass{article}
\usepackage[sumlimits,intlimits,namelimits]{amsmath}
\usepackage{amssymb}
\usepackage[english]{babel}
\usepackage{bbm}
\usepackage{parskip}

\newcommand{\I}{\text{Im}\;}

\newcommand{\be}{\begin{eqnarray}}
\newcommand{\ee}{\end{eqnarray}}

\newcommand{\lb}{\label}
\def\>{\rangle}
\def\<{\langle}

\begin{document}
%
\begin{titlepage}
\hfill{LMU-ASC 35/08}
\vspace*{.5cm}
\begin{center}
{\large{{\bf Quasi-Normal Modes in Topologically Massive Gravity}}} \\
\vspace*{1.5cm}
Ivo Sachs\footnote{{\tt
Ivo.Sachs@physik.uni-muenchen.de}}\\
{\it Arnold Sommerfeld Center for Theoretical Physics (ASC),
Ludwig-Maximilians Universit\"{a}t\\
Theresienstrasse 37\\
D-80333, M\"{u}nchen\\
Germany}\\
\vspace*{.5cm} Sergey N. Solodukhin\footnote{ {\tt
solodukh@lmpt.univ-tours.fr}}
\\
\it{Laboratoire de Math\'ematiques et Physique
Th\'eorique CNRS-UMR 6083, \\
Universit\'e de Tours, Parc de Grandmont, 37200 Tours, France\\
and CERN Theory Division, CH 1211 Geneva 23, Switzerland}
\end{center}
\vspace*{1cm}
\begin{abstract}
\noindent {
We determine
the black hole quasi-normal mode spectrum for tensor perturbations in topologically massive  AdS-gravity. In the special case of chiral gravity quasi-normal
modes are absent despite of the presence of a horizon. In the process we uncover a simple algebraic structure
in the quasi normal modes spectrum: the tower of QNM's consists of descendents of a "chiral highest weight'' QNM which
in turn satisfies a first order equation.
\\
\vspace*{.25cm}
}
\end{abstract}
\vspace*{.25cm}
\end{titlepage}
\section{Introduction}
The question whether and in what sense 3-dimensional gravity with
a negative cosmological constant makes sense as a unitary quantum
theory has a long history. Since there are no propagating bulk
degrees of freedom the dynamics appears to be entirely contained
in the boundary degrees of freedom. This was made precise in the
re-formulation of 3-dimensional gravity as a pair of Chern-Simons
Lagrangians \cite{Townsend,Witten98} which was subsequently
reduced to a Liouville theory on the boundary of $AdS_3$
\cite{9405171,9506019}. A neccessary condition for the viability
of any quantum theory of $AdS_3$ gravity is the requirement that
the semiclassical properties of  the 3-dimensional BTZ-black holes
\cite{9204099} are correctly encoded in the microscopic theory.
While the semi-classical Hawking radiation \cite{9409047} is
successfully reproduced in this formulation
\cite{Birmingham:1997rj,ES}, Liouville theory fails to reproduce
the entropy of the 3-dimensional BTZ-black holes \cite{9806026}.
This inconsistency can be removed by embedding $AdS_3$-gravity
into super string theory on $AdS_3\times S^3\times M^4$
\cite{9601029} which is dual to $Q_5$ D5 - and $Q_1$ D1 brane
system on $M^4\times S^1$. This is a large $N$ duality in the
sense that the supergravity approximation is valid for $g_s\!\to\!
0$, $N\!=\!\sqrt{Q_1Q_5} \!\to\! \infty$. Apart from enlarging the
field content of $AdS_3$-gravity to the full supergravity
multiplet, one is faced with the question of the validity of this
duality at finite $N\gg 1$.

It was already noticed in \cite{0106112} that the retarded 2-pt
function in a BTZ black hole background decays exponentially in
time whereas at finite N the dual CFT is Poincar\' e-recurrent
\cite{BSS2}. A possible resolution may be to replace the SUGRA
background by a weighted sum over all asymptoically $AdS$
geometries as in \cite{0712.0573}. However,  unless the
contributions from geometries with horizon has zero measure,
Poincar\'e recurrence cannot possibly be reproduced on the gravity
side \cite{BSS2}. Another logical possibility is to replace the
SUGRA background by a sum over micro-state geometries  as
suggested in \cite{0109154}. The BTZ horizon should then appear as
a result of corse graining.

Recently, arguments were given in \cite{Witten07} in favor of a
duality between pure gravity on $AdS_3$ and  a holomorphically
factorized CFT. Now holomorphic factorization implies Poincar\'e
recurrence trivially. On the other hand, asymptotically $AdS$
geometries with a horizon contribute to the partition function of
the dual CFT. Since pure Einstein gravity with a negative
cosmological constant has no propagating degrees of freedom the
presence of a horizon does not need to contradict holomorphic
factorization. Of course, a scalar field does propagate on the
black hole back ground and the corresponding quasi-normal modes
imply exponential decay of the real-time correlator. But a scalar
field is not a degree of freedom in pure gravity. On the other
hand, Einstein gravity in  3-dimensions can be supplemented with a
gravitational Chern-Simon term \cite{Deser:1982vy}. In that case
\cite{Witten07} produces evidence that for integer values of the
cosmological constant this theory may be dual to a pair of
left/right CFT's with different central charges. The difference is
that the topologically massive theory obtained in this way has a
propagating graviton so that we expect generically that the
presence of a horizon will spoil holomorphic factorization of the
dual CFT at finite temperature.

In the present paper we analyse the structure of gravitational
quasi-normal modes for topologically massive gravity. We find a
infinite set of  quasi-normal modes in the BTZ-backgroud  for
generic values of the CS-coupling thus indicating the exponential
decay of the retarded correlation function of the energy-momentum
tensor in the "dual CFT". The case of chiral gravity is special.
Although there is a BTZ-black hole in chiral topological gravity
there is no gravitational quasi-normal mode associated with it.
This is then consistent with the proposal that this theory is dual
to chiral CFT on the boundary. For generic values of the
CS-coupling, however it appears that the only way we can possibly
recover  Poincar\'e recurrence  on the gravity side is by summing
over a continuum of non-geometrical configurations, possibly by
treating the spin-connection and the 3-bein as independent
variables, in such a way that the geometric contributions with
horizon have zero measure.

\section{Preliminaries}
We take as the starting point the action
 \be
S&=&-\frac{1}{\kappa^2}\int\left(R+\frac{2}{l^2}\right)+\frac{1}{4m
\kappa^2}\int\left(\epsilon^{\mu\nu\alpha} R_{\mu\nu ab}\omega_{\
,\alpha}^{ab}+\frac{2}{3}\omega^a_{\ b,\mu}\omega^b_{\
c,\nu}\omega^c_{\ a,\alpha}\right) \lb{action}
\ee
where
$\kappa^2=16\pi G$. The parameter $l$ sets the AdS scale, in what
follows we will use units with $l=1$.

The parameter $m$ is related to the levels $k_{L/R}$ in the
Chern-Simons formulation of 3d gravity \cite{Witten07} through
m=$(k_R+k_L)/(k_R-k_L)$. In particular, if one assumes that the
quantization condition for $k$ in CS-theory applies, then $m$ is a
rational number with $|m|\geq 1$. The action (\ref{action}) leads
to the equation of motion \be (1)_{\mu\nu}+(2)_{\mu\nu}=0~~, \ee
where
\be
(1)_{\mu\nu}&=&R_{\mu\nu}-\frac{1}{2}g_{\mu\nu}R- g_{\mu\nu}\\
(2)_{\mu\nu}&=&\frac{1}{m}\epsilon_{\mu}^{\,\,\alpha\lambda}\nabla_\alpha
(R_{\lambda\nu}-\frac{1}{4}g_{\lambda\nu}R) \nonumber
\ee Notice, that the solution to these equations
is space-time with constant scalar curvature, $R=-6$. The equation
for the linear metric perturbations is then given by the variation
of the equation of motion. This gives \be \delta
(1)_{\mu\nu}&=&\frac{1}{2}\nabla_\kappa\nabla_\nu\,
h^\kappa_{\,\mu} + \frac{1}{2} \nabla_\kappa\nabla_\mu\,
h^\kappa_{\,\nu}- \frac{1}{2} \nabla^2 h_{\mu\nu}- \frac{1}{2}
\nabla_\mu\nabla_\nu\, h^\kappa_\kappa+2 h_{\mu\nu}\nonumber \\
&&-\frac{1}{2} g_{\mu\nu}\nabla^\kappa\nabla^\gamma\,
h_{\kappa\gamma}+   \frac{1}{2} g_{\mu\nu} \nabla^2
h^\kappa_{\;\kappa}-
 g_{\mu\nu} h^\kappa_{\;\kappa}\\
\delta
(2)_{\mu\nu}&=&\frac{1}{4m}\left[\epsilon_{\mu}^{\,\,\alpha\lambda}\nabla_\alpha
h_{\lambda\nu} +\epsilon_{\nu}^{\,\,\alpha\lambda}\nabla_\alpha
h_{\lambda\mu}\right]\nonumber \\
&&+\frac{1}{4m}\left[\epsilon_{\mu}^{\,\,\alpha\lambda}\nabla_\alpha\nabla_\kappa\nabla_\nu
h^\kappa_{\,\lambda}   +
\epsilon_{\nu}^{\,\,\alpha\lambda}\nabla_\alpha\nabla_\kappa\nabla_\mu
h^\kappa_{\,\lambda} \right]\nonumber \\
&&-\frac{1}{4m}\left[\epsilon_{\mu}^{\,\,\alpha\lambda}\nabla_\alpha\nabla^2
h_{\lambda\nu}+\epsilon_{\nu}^{\,\,\alpha\lambda}\nabla_\alpha
\nabla^2h_{\lambda\mu}\right]\nonumber \lb{equations} \ee To
continue we need to fix the gauge. We impose \be \nabla^\mu
h_{\mu\nu}=0~,~~g^{\mu\nu}h_{\mu\nu}=0~~. \lb{gauge} \ee With this
choice the equations
$$
\delta(1)_{ \mu\nu}+\delta(2)_{\mu\nu}=0
$$
for the linear perturbations on the background of locally AdS
metric with the Riemann curvature $R^{\mu\nu}_{\ \
\alpha\beta}=-(\delta^\mu_\alpha
\delta^\nu_\beta-\delta^\mu_\beta\delta^\nu_\alpha)$,
 take a simple form \be
(\nabla^2+2)h_{\mu\nu}+{1\over m}\epsilon_\mu^{\
\alpha\beta}\nabla_\alpha (\nabla^2+2)h_{\beta \nu}=0~~\lb{lin}
\ee or equivalently, since the two differential operators in
(\ref{lin}) commute, \be (\nabla^2+2)\left[\epsilon_\mu^{\
\alpha\beta}\nabla_\alpha h_{\beta \nu} +mh_{\mu\nu}\right] =0~~.
\lb{lin2} \ee
 The equations of motion for the linear perturbations
thus split on two equations: the equation for a  massless graviton
$(\nabla^2+2)h_{\mu\nu}=0$ which does not have propagating
solutions and the first order equation for a massive graviton
\begin{equation}\label{first}
\epsilon_\mu^{\;\;\alpha\beta}\nabla_\alpha h_{\beta\nu}+m
h_{\mu\nu}=0~~.
\end{equation}
Notice that the gauge conditions (\ref{gauge}) follow
automatically from (\ref{first}) which thus contains information
about both the field equations and the gauge. The parameter $m$
introduced as coupling in the gravitational action (\ref{action})
thus has the meaning of the mass of the graviton. It is interesting that
the originally higher derivative equations of motion reduce to a
first order in derivative equation (\ref{first}). This fact seems
to be rather general \cite{Tyutin:1997yn} and is valid for any
massive field of spin $s \geq 1$ in three-fimensional AdS
background.

Squaring the equation (\ref{first}) produces the second order
equation \be (\nabla^2+3-m^2)h_{\mu\nu}=0~~. \lb{massive} \ee In
what follows we will, however, work directly with the first order
equation (\ref{first}) in order to avoid ambiguities with the sign
of $m$ present in  (\ref{massive}).

In the AdS/CFT correspondence the bulk theory of  topologically
massive gravity is equivalent to a conformal field theory with
brocken local Lorentz invariance. This theory is characterized by
different left and right central charges \cite{Kraus:2005zm,Solodukhin:2005ah} \be c_L={3\over 2G}(1+{1\over
m})~,~~c_R={3\over 2 G}(1-{1\over m})~~. \lb{central} \ee The
values $m=1$ and $m=-1$ are special in that one of the central
charges vanishes and the dual CFT contains only one chiral sector.

The change in the Riemann tensor produced by the metric perturbation
(over AdS background) that satisfies equation (\ref{massive}) is
\be \delta R^{\mu\nu}_{\ \ \alpha\beta}={(1-m^2)\over
2}(h^\mu_\alpha \delta^\nu_\beta+h^\nu_\beta
\delta^\mu_\alpha-h^\mu_\beta
\delta^\nu_\alpha-h^\nu_\alpha\delta^\mu_\beta ) \lb{change} \ee
so that the perturbation with $m^2=1$ produces no change in the
curvature and thus, locally, is a pure gauge (see also
\cite{Strominger}, \cite{1}, \cite{2} and \cite{Park:2008yy} for a related
discussion).

\section{Quasi-Normal Modes}

\subsection{Local $SL(2,R)\times SL(2,R)$ algebra of Killing vector fields}
The black hole solutions of Einstein Gravity with a negative
cosmological constant are also solutions in topologically massive
gravity.  Here we consider the non-rotating\footnote{Rotation can
be included by a suitable change of coordinates.} black hole with
unit mass given by the metric
\begin{equation}
ds^2=-\sinh^2(\rho) d\tau^2+\cosh^2(\rho) d\phi^2 +d\rho^2~~.
\end{equation}
In what follows we will work in light cone coordinates $u=\tau+\phi$, $v=\tau-\phi$, in which
\begin{equation}\label{guv}
g=\begin{pmatrix} \frac{1}{4}&-\frac{1}{4}\cosh(2\rho)&0\cr
        -\frac{1}{4}\cosh(2\rho)&\frac{1}{4}&0\cr
        0&0&1
        \end{pmatrix}~~.
\end{equation}
The black hole metric (\ref{guv}) admits two sets of the  Killing vector fields, $L_k$ and $\bar L_{k}$, $k=0,-1,1$ defined as
\begin{eqnarray}
L_0&=&-\partial_u\nonumber\\
L_{-1}&=&e^{-u}\left(-\frac{\cosh(2\rho)}{\sinh(2\rho)}\partial_u-\frac{1}{\sinh(2\rho)}
\partial_v-\frac{1}{2}\partial_\rho\right)\lb{Kf}\\
L_{1}&=&e^{u}\left(-\frac{\cosh(2\rho)}{\sinh(2\rho)}\partial_u-\frac{1}{\sinh(2\rho)}\partial_v+
\frac{1}{2}\partial_\rho\right)\nonumber
\end{eqnarray}
and similarly for $\bar L_0,\bar L_{1},\bar L_{-1}$ defined as
(\ref{Kf}) by substituting $u\rightarrow v$ and $v\rightarrow u$.
Locally they form  a basis of the Lie algebra $SL(2,R)$,
\be\label{alg} [L_0, L_{\pm 1} ]=\mp L_{\pm
1}~,~~[L_1,L_{-1}]=2L_0~~.  \ee
Note that these vector fields
cannot be integrated to generate global symmetries of the black
hole background since they do not commute with the (hyperbolic)
elements of $SL(2,R)$ which provide the discrete identifications
of points in $AdS_3$ required to produce the black hole
(\ref{guv}). In particular, they do not preserve the boundary
conditions when acting on functions defined on the black hole
manifold. Nevertheless  they turn out to be quite useful since, as
we will now show, they generate the whole tower of quasi normal
modes.

The vector fields $L_k$ and $\bar L_k$ satisfy the usual Killing
equation $\nabla_\mu L_\nu+\nabla_\nu L_\mu=0$. In three
dimensional locally AdS spacetime chacterised by the Riemann
tensor $R^\alpha_{\ \beta\mu\nu}=-(\delta^\alpha_\mu
g_{\beta\nu}-\delta^\alpha_\nu g_{\beta\mu})$ the Killing vectors
satisfy in addition the following second order equations
$$
\nabla_{(\mu}\nabla_{\nu)} L_\alpha=g_{\mu\nu}L_\alpha-{1\over
2}(L_\mu g_{\nu\alpha}+L_\nu g_{\mu\alpha})~~,
$$
$$
\nabla_{[\mu}\nabla_{\nu ]}L_\alpha={1\over 2}(L_\mu
g_{\nu\alpha}-L_\nu g_{\mu\alpha})~~.
$$
Using these equations one shows that the Lie derivatives with
respect to the Killing vectors $L_k$ are compatible with the zero
trace, transversality condition (\ref{gauge})  on $h_{\mu\nu}$, i.e. (\ref{gauge}) implies
\be g^{\mu\nu} (L_{k}h_{\mu\nu})=0~,~~\nabla^\mu
(L_{k}h_{\mu\nu})=0~,~~k=0,-1,1\,  .\lb{Lie} \ee
Similarly the
equation of motion (\ref{first})  implies \be
\epsilon_\mu^{\;\;\alpha\beta}\nabla_\alpha (L_k h)_{\beta\nu}+m (L_k
h)_{\mu\nu}=0~~. \ee

\subsection{Scalar Modes}
Before constructing the gravitational quasi-normal modes, as a
warm-up we will reproduce the scalar quasi-normal modes
\cite{Birmingham:2001hc,BSS1} in the BTZ background using the algebra of the Killing
fields (\ref{alg}).  The scalar field equation can be written in
the form \be\label{ns} (-\nabla^2+m_s^2)\varphi=\left[2(L^2+\bar
L^2)+m_s^2 \right]\varphi=0~~, \ee where $L^2={1\over 2}(L_1
L_{-1}+L_{-1}L_1)-L^2_0$ is the Casimir operator of $SL(2,R)$. In
the process we will uncover a new structure present in the set of
quasi-normal modes. In particular we will see that rather similar
to the "highest weight" solutions in $AdS_3$ found in
\cite{Maldacena:1998bw}, the quasi-normal modes are descendents of a
solution to a first order differential equation. The difference,
however, is that only a half of the ``highest weight'' conditions
should be imposed.

To see this we make the Ansatz
\begin{equation}
 \Phi=e^{-ip_+u-ip_- v}F(\rho)
\end{equation}
so that $L_0\Phi=ip_+\Phi$ and $\bar L_0\Phi=ip_- \Phi$, note that
$\omega=p_++p_-$ and $k=p_+-p_-$ are respectively the energy and
the $\phi$-momentum. Following \cite{Maldacena:1998bw} one could impose
the ``highest weight'' conditions $L_{1}\Phi=\bar L_{1}\Phi=0$
reflecting the fact that there are two $SL(2,R)$ algebras locally
present in the geometry. These conditions are perfectly suited to
reproduce the so-called ``normalizable modes'' in anti-de Sitter
space-time studied in \cite{Balasubramanian:1998sn} but they
appear to be too strong in the black hole background because the
descendents of such highest weight modes have imaginary
$\phi$-momentum.

This problem can be avoided, however, by imposing a weaker condition which allows for real $\phi$-momentum, namely
\be\label{LP}
 L_{1}\Phi=0~~.
\ee
This is exactly a half of the ``highest weight'' condition. Therefore, we call it a ``chiral highest weight''
condition.
 This condition then implies a first order equation for $F(\rho)$
\begin{equation}
\left(2i p_+\cosh(2\rho)+2i p_-+\sinh(2\rho)\partial_\rho\right)F(\rho)=0~~.
\end{equation}
If one were to impose in addition $\bar L_1\phi=0$, then it is clear from  (\ref{alg}) that the the wave equation  (\ref{ns})  reduces to an algebraic equation for $p_\pm$. It is not obvious that this should still hold when imposing just one condition $L_1\Phi=0$. However, it turns out that this is indeed the case.
The key observation is that
\be
\bar L_1\bar L_{-1}\Phi=(p_+^2-p_-^2+i(p_++p_-))\Phi ~~.
\ee
With this the scalar wave equation (\ref{ns})
reduces to the algebraic equation
\begin{equation}
4p_+^2+4i p_++m_s^2=0 \lb{al}~~.
\end{equation}
Notice that only $p_+$ but not $p_-$ appears in (\ref{al}). Thus
\begin{equation}\label{sqnm1}
 p_+=-i h_L(m_s)~,~~ h_L(m_s)={1+\sqrt{1+m_s^2}\over 2}\ \,.
\end{equation}
If the BTZ black hole is interpreted as part of the gravitational dual to the $N=(4,4)$ superconformal field theory on the boundary, then  $h(m_s)$ is the (left moving) conformal weight of the dual operator in the CFT on the boundary \cite{BSS1}.
Reality of the $\phi$-momentum then requires $p_+-p_-=k$, where $k$ is real but otherwise unconstraint.
If the direction along the coordinate $\phi$ is compact as it is for the BTZ case the momentum $k$ along this direction should be integer.
The solution is then given by
\begin{eqnarray}\label{SQNML}
\Phi&=&e^{-2h_L(m_s)\tau+ik(\tau-\phi)}F(\rho)\\
F(\rho)&=&(\sinh(\rho))^{-2h_L(m_s)}(\tanh(\rho))^{ik}~~. \lb{sol}
\end{eqnarray}
This solution is ingoing at the horizon, falls off in time as well as at infinity. The corresponding frequency
\be\label{pole}
\omega^L=-k-2ih_L(m_s)
\ee
is the lowest quasi-normal mode in the left-moving sector.
Note that we define the left- and right-moving sectors from the boundary CFT point of view.

The quasi-normal mode (\ref{SQNML}) determines the location of the
pole, in the lower half plane, nearest to the real axis, of the
retarded correlation function of a left-moving chiral operator,
$V_{h_L}(\tau+\phi)$, in the boundary CFT with conformal weight
$h_L(m)$ \cite{BSS1}. Note that the bulk quasi-normal mode
corresponding to this pole is right moving.

For negative $m_s^2$ such that $-1\leq m_s^2<0$ there exists second solution to the
equation (\ref{alg}) corresponding to the conformal weight $h_L={1-\sqrt{1+m_s^2}\over 2}$
that would produce a second set of quasi-normal modes.
This is consistent with the observation made in \cite{BSS1}.

Furthermore, provided $h_L(m_s)\neq 0$ then, acting by $L_{-1}\bar
L_{-1}$ on  (\ref{SQNML}) produces again a quasi-normal mode with
the same asymptotic fall-off behaviour as  (\ref{SQNML}) but  with
$\I\omega\to \I\omega -2$. Thus, the condition  $ L_{1}\Phi=0$
together with an algebraic condition on  $p_+$ and $p_-$ leads to
an infinite tower of quasi normal scalar modes \be
\Phi^{(n)}=(L_{-1}\bar{L}_{-1})^n\Phi \ee that are descendents of
the mode $\Phi$. In particular, the asymptotic fall-off behavior
of $\Phi^{(n)}$ for large $\rho$ is uniquely determined by
$h_L(m_s)$ and is independent of $n$.

The corresponding frequencies \be
\omega^L_n=-k-2i(h_L(m_s)+n)~,~~n\in N \lb{tower} \ee are exactly
the quasi-normal frequences in this chiral sector found in
\cite{BSS1,Birmingham:2001hc}. For $h_L(m_s)=0$   solution (\ref{SQNML}) is not a
quasi-normal mode since it does not satisfy the quasi-normal
boundary condition.

If instead of (\ref{LP}) we impose $ \bar L_{1}\Phi=0$ this leads to the first order equation for $F(\rho)$
\begin{equation}
\left(2i p_-\cosh(2\rho)+2i  p_++\sinh(2\rho)\partial_\rho\right)F(\rho)=0~~.
\end{equation}
Imposing the field equation the gives
\begin{equation}\label{sqnm1R}
 p_-=-ih_R(m_s)~~,
\end{equation}
where $ h_R(m_s)=h_L(m_s)$ for a scalar perturbation. The
corresponding quasi-normal modes can be obtained form the previous
one by the substitution $p_+\leftrightarrow p_-$ which results in
$\tau-\phi \to \tau+\phi$ in (\ref{sol}) and thus reproduces the
right-moving copy of the infinite tower (\ref{tower}). Again,
repeated action by  $L_{-1}\bar L_{-1}$ produces the whole tower
of  right moving quasi-normal modes, \be
\omega^R_n=k-2i(h_R(m_s)+n)~,~~n\in N \lb{towerR} \ee in complete
agreement with \cite{BSS1}.

To summarize, we find that the complete set of scalar quasi normal
modes in the BTZ background \cite{BSS1} is generated starting from
the  "chiral highest weight" conditions  $ L_{1}\Phi=0$ or  $ \bar
L_{1}\Phi=0$ . The remaining possibilities  $ \bar L_{-1}\Phi=0$
and  $ L_{-1}\Phi=0$ do not lead to quasi-normal modes since the
solutions to this equation have outgoing flux at the horizon.

\subsection{Tensor Modes}
Let us now turn to the gravitational quasi-normal modes for
topologically massive gravity. As explained in section 2, in the
gauge (\ref{gauge}) the equation for the massive graviton
satisfies a first order equation of motion (\ref{first}),
\begin{equation}\label{33}
\epsilon_\mu^{\;\;\alpha\beta}\nabla_\alpha h_{\beta\nu}+m
h_{\mu\nu}=0~~.
\end{equation}

We make the Ansatz \cite{Strominger} (in $u,v,\rho$ coordinates)
\begin{equation}\label{AnsatzB}
h_{\mu\nu}=e^{-ip_+u-ip_-v}\psi_{\mu\nu}(\rho)~,~~p_++p_-=\omega~,~~p_+-p_-=k
\end{equation}
for the metric perturbation where
\begin{equation}
\psi_{\mu\nu}=F(\rho)\begin{pmatrix} 1&0&\frac{2}{\sinh(2\rho)}\cr
        0&0&0\cr
        \frac{2}{\sinh(2\rho)}&0&\frac{4}{\sinh^2(2\rho)}
        \end{pmatrix}~~.
\end{equation}
The dominant component of $h_{\mu\nu}$ at infinity is $h_{uu}$. The transversality condition  $\nabla_\mu h^\mu_{\;\;\nu}=0$ then implies
\begin{equation}\label{tB}
\left(2ip_++2ip_-\cosh(2\rho)+
\sinh(2\rho)\partial_\rho\right)F(\rho)=0~~.
\end{equation}
A slightly tedious, but straight forward calculation shows that for our Ansatz the transversality condition  is,
in fact, equivalent to the ``chiral highest weight'' condition\footnote{For a generic symmetric tensor $t_{\mu\nu}$
the transversality condition (\ref{tB})
can, of course, not be equivalent to  $\bar L_1 t_{\mu\nu}=0$ since the latter impose six constraint rather than three in (\ref{tB}).}
 \be
\bar L_1 h_{\mu\nu}=0~~.
\ee
In particular, equation (\ref{tB}) takes the form  of the equation $\bar L_1 F=0$ for the scalar field $F$.

In addition we need to satisfy the first order equation of motion (\ref{first}).
Using $\epsilon^{\rho u v}={1\over \sqrt{-g}}={4\over \sinh 2\rho}$ we get from (\ref{first}) for $\mu=\nu=\rho$ that
\begin{eqnarray}\label{cond}
p_-=-ih_R(m)\;\;,\;\;h_R(m)=-\frac{1}{2}-\frac{m}{2}
\end{eqnarray}
where now $h_R(m)$ is the right-moving conformal weight  (for
negative $m$). For $m=-1$, the weight $h_R(m)$ vanishes. If
condition (\ref{cond}) is satisfied then the remaining equations
in (\ref{first}) reduce to the transversality condition
(\ref{gauge}). It is clear that (\ref{first}) implies
(\ref{gauge}) for any symmetric tensor. What we have just shown is
that the converse holds true for the Ansatz (\ref{AnsatzB})
provided  (\ref{cond}) is satisfied. The momentum in
$\phi$-direction $k=p_+-p_-$ should be integer if $\phi$ is
compact. We thus get
\begin{equation}\label{bsol1}
h_{\mu\nu}=e^{-2h_R(m)\tau-ik(\tau+\phi)}\psi_{\mu\nu}(\rho)~~.
\end{equation}
The solution  (\ref{tB}) is then given by
\begin{equation}\label{solBF}
F(\rho)=(\sinh(\rho))^{-2h_R(m)}(\tanh(\rho))^{-ik}
\end{equation}
and thus the solution is completely determined. For $m<-1$ this
solution is ingoing at the horizon and falls off at infinity as
well as in time. It is thus a genuine quasi-normal mode. The
Ansatz (\ref{AnsatzB}) for the metric perturbation together with
the asymptotic behavior  of (\ref{solBF}) has the right asymptotic
form in order to couple to the right moving component of a
two-index tensor $T_{vv}$ in the boundary CFT since the $v\rho$
and the $\rho\rho$ components are subdominant at infinity. In what
follows we will require this asymptotic structure for quasi-normal
mode solutions.

In analogy with scalar modes we can act on the ``chiral highest
weight'' quasi-normal mode with $L_{-1}\bar L_{-1}$. The effect of
this will be to replace $\I\omega\to \I\omega -2$ in
(\ref{bsol1}). As was discussed in section 3.1,
$$
h^{(n)}_{\mu\nu}=(L_{-1}\bar L_{-1})^nh_{\mu\nu}
$$
is  transverse and traceless. Furthermore, $L_{-1}\bar L_{-1}$
commutes with the equation of motion (\ref{first}). Thus
$h^{(n)}_{\mu\nu}$ is again a solution of the equation of motion
with the same asymptotic fall-off behavior as (\ref{bsol1}).
Consequently, the complete tower of right-moving
gravitational quasi-normal modes is generated in this way from the
basic solution (\ref{bsol1}). The corresponding quasinormal
frequencies are given by
\be \omega^R_n=k-2i(h_R(m)+n)~, ~~n\in N
\lb{om} \ee

For chiral gravity $m=-1$ the solution (\ref{bsol1}) has constant amplitude (in $\rho$ and $t$)
and is thus not quasi-normal.

We can obtain a second solution by choosing
\begin{equation}
\psi_{\mu\nu}=F(\rho)\begin{pmatrix} 0&0&0\cr
        0&1&\frac{2}{\sinh(2\rho)}\cr
        0&\frac{2}{\sinh(2\rho)}&\frac{4}{\sinh^2(2\rho)}
        \end{pmatrix}~~.
\end{equation}
In this case we have
\begin{equation}
\nabla_\mu h^\mu_{\;\;\nu}=0\implies\left(2ip_-+2i p_+\cosh(2\rho)+\sinh(2\rho)\partial_\rho\right)F(\rho)=0
\end{equation}
which, in turn, is equivalent to $ L_1 h_{\mu\nu}=0$. Furthermore,
the Ansatz  (\ref{AnsatzB}) satisfies the first order equation of motion (\ref{first}) provided
\begin{eqnarray}
  p_+=-ih_L(m)\;\;,\;\;\;h_L(m)=\frac{m}{2}-\frac{1}{2}~~.
\end{eqnarray}
Since $p_+-p_-=k$ we then get
\begin{equation}\label{bsol2}
h_{\mu\nu}=e^{-2h_L(m)\tau+ik(\tau-\phi)}\psi_{\mu\nu}(\rho)
\end{equation}
and
\begin{equation}\label{solBF2}
F(\rho)=(\sinh(\rho))^{-2h_L(m)}(\tanh(\rho))^{ik}~~.
\end{equation}
This is the basic quasi-normal mode for $m>1$. It couples to the
left-moving components $T_{uu}$ of a 2-tensor. The higher
quasi-normal modes are again obtained by acting with  $L_{-1}\bar
L_{-1}$. The corresponding frequencies are \be
\omega^L_n=-k-2i(h_L(m)+n)~,~~n\in N \lb{l} \ee For chiral gravity
$m=1$ the quasi-normal mode is absent. In analogy with the scalar
modes the conditions $ L_{-1} h_{\mu\nu}=0$ and $ \bar L_{-1}
h_{\mu\nu}=0$ lead to unphysical solutions with outgoing flux at
the horizon.

\section{Conclusion}
In this paper we have found the quasi-normal mode spectrum for
black holes in topologically massive gravity. We found that the
quasi-normal modes satisfy a first order chiral highest weight
equation. While the presence of a first order equation for
gravitons can be understood from the form of the action, the fact
that this equation agrees with a highest weight condition of  the
$SL(2,R)$ algebra suggests that the highest weight condition
should be understood as the implementation of the chirality
condition on the gravitational excitations, ie. up to the
exponential decay in time they only depend on $u$ or $v$. Since the
gravitational excitations are chiral as consequence of the CS-term
the equation of motion and the highest weight condition should
coincide. That the same structure appears for the scalar QNM's is less obvious since they satisfy a second order differential equation.

We should also mention  that for a given sign of $m$ there are
only quasi-normal modes with definite chirality. Form the
gravitational point of view it is clear that this has to be so
because the CS-term in the action is odd under the exchange of $u$
and $v$. Thus given a solution $h_{\mu\nu}(\tau,v)$ the
corresponding function $h_{\mu\nu}(\tau,u)$ is then necessarily a
solution of the equation of motion of the action with the opposite
sign of $m$. On the CFT side this leads to the interesting
prediction that  for $m<0$ the retarded Green function of the
right moving primary $T_{uu}$, if it exists, has no poles in the
lower half complex frequency plane which should then imply that
the real-time Green function is quasi-periodic even at finite
temperature. Similarly for $T_{vv}$ and $m>0$.

For chiral gravity ($m=1$ or $m=-1$) we did not find any
quasi-normal modes. This may not be surprising since for these
values of $m$ the tensor perturbations (\ref{33}) satisfy the wave equation
for massless gravitons which are known not to propagate. It should
be noted that we only analysed solutions to the first order
equation (\ref{first}). There may be however solutions to the
third order in derivative equation (\ref{lin2}) that are not
solutions to (\ref{first}) or the massless graviton equation. In
recent work \cite{2} it was found that for $m^2=1$  equation
(\ref{lin2}) has a non-trivial solution which grows  at the
boundary. This solution, however, is not a quasi normal mode since it does not satisfy the QNM boundary conditions and grows in time\footnote{ We thank
D. Grumiller for raising this point.}.

\subsection*{Acknowledgments:}
I.S. an S.S. were supported in parts by the Transregio TRR~33 `The Dark
Universe'. In addition  I.S. acknowledges support from the Excellence Cluster `Origin and Structure of the
Universe' of the DFG as well as the DFG grant  Ma 2322/3-1. S.S. thanks  X. Bekaert for useful
discussions and R. Metsaev for pointing out reference \cite{Tyutin:1997yn}. S.S. is grateful to the Theory Division at CERN for the hospitality extended to him while this work was completed.


\end{document}